\documentclass[prx,twocolumn,byrevtex,floatfix,superscriptaddress,longbibliography]{revtex4-2}

\usepackage{gensymb}
\usepackage{amsmath}    % need for subequations
\usepackage{amsfonts}
\usepackage{graphicx}   % need for figures
\usepackage{verbatim}   % useful for program listings
\usepackage{color}      % use if color is used in text
\usepackage{subfigure}  % use for side-by-side figures
\usepackage{hyperref}   % use for hypertext links, including those to external documents and URLs
\usepackage{bm}			% bold math
\begin{document}

\title{\boldmath Vital role of anisotropy in cubic chiral skyrmion hosts \unboldmath}

\author {M.~Prei{\ss}inger}
\affiliation{Experimentalphysik V, Center for Electronic Correlations and Magnetism, Institute of Physics, University of Augsburg, D-86135 Augsburg, Germany}

\author{K.~Karube}
\affiliation{RIKEN Center of Emergent Matter Science (CEMS), Wako 351-0198, Japan}

\author{D.~Ehlers}
\affiliation{Experimentalphysik V, Center for Electronic Correlations and Magnetism, Institute of Physics, University of Augsburg, D-86135 Augsburg, Germany}

\author{B.~Szigeti}
\affiliation{Experimentalphysik V, Center for Electronic Correlations and Magnetism, Institute of Physics, University of Augsburg, D-86135 Augsburg, Germany}

\author{H.-A.~Krug~von~Nidda}
\affiliation{Experimentalphysik V, Center for Electronic Correlations and Magnetism, Institute of Physics, University of Augsburg, D-86135 Augsburg, Germany}

\author{J.~S.~White}
\affiliation{Laboratory for Neutron Scattering and Imaging, Paul Scherrer Institute, CH-5232 Villigen, Switzerland}

\author{V.~Ukleev}
\affiliation{Laboratory for Neutron Scattering and Imaging, Paul Scherrer Institute, CH-5232 Villigen, Switzerland}

\author{H.~M.~R{\o}nnow}
\affiliation{Laboratory for Quantum Magnetism, Institute of Physics, \'Ecole Polytechnique F\'ed\'erale de Lausanne, CH-1015 Lausanne, Switzerland}

\author{Y.~Tokunaga}
\affiliation{Department of Advanced Materials Science, University of Tokyo, Kashiwa 277-8561, Japan}

\author{A.~Kikkawa}
\affiliation{RIKEN Center of Emergent Matter Science (CEMS), Wako 351-0198, Japan}

\author{Y.~Tokura}
\affiliation{RIKEN Center of Emergent Matter Science (CEMS), Wako 351-0198, Japan}
\affiliation{Tokyo College and Department of Applied Physics, University of Tokyo, Bunkyo-ku 113-8656, Japan}

\author{Y.~Taguchi}
\affiliation{RIKEN Center of Emergent Matter Science (CEMS), Wako 351-0198, Japan}

\author{I.~K\'ezsm\'arki}
\affiliation{Experimentalphysik V, Center for Electronic Correlations and Magnetism, Institute of Physics, University of Augsburg, D-86135 Augsburg, Germany}

\date{\today}% It is always \today, today,

\begin{abstract}
The impact of magnetic anisotropy on the skyrmion lattice (SkL) state in cubic chiral magnets has been overlooked for long, partly because a semi-quantitative description of the thermodynamically stable SkL phase pocket forming near the Curie temperature could be achieved without invoking anisotropy effects. However, there has been a range of phenomena reported recently in these materials, such as the formation of low-temperature tilted conical and SkL states as well as temperature-induced transformations of lattice geometry in metastable SkL states, where anisotropy was suspected to play a key role. To settle this issue on experimental basis, we quantified the cubic anisotropy in a series of CoZnMn-type cubic chiral magnets. We found that the strength of anisotropy is highly enhanced towards low temperatures in these compounds, moreover, not only the magnitude but also the character of cubic anisotropy drastically varies upon changing the Co/Mn ratio. We correlate these changes with temperature- and composition-induced variations of the helical modulation vectors, the anharmonicity and structural rearrangements of the metastable SkLs and the spin relaxation rates. Similar systematic studies on magnetic anisotropy may not only pave the way for a quantitative and unified description of the stable and metastable modulated spin textures in cubic chiral magnets but would also help exploring further topological spin textures in this large class of skyrmion hosts.
\end{abstract}

\maketitle

\section{introduction}

In recent years, magnetic skyrmions have been explored in a great variety of materials, not only in single crystals but in ultrathin films as well as hetero- and nanostructures~\cite{Heinze2011, Wilson2012, Yu2013, Du2015, Luchaire2016, Stolt2017, Herve2018, Mathur2019, Stolt2019}. Periodic 2D lattices of Bloch skyrmions have been first observed in cubic chiral magnets~\cite{Muehlbauer2009, Yu2010, Muenzer2010, Yu2011, Kanazawa2012, Seki2012, Milde2013, Tanigaki2015, Qian2016, Pappas2017}. Already in 1976, the archetypical cubic chiral magnet MnSi, in which the formation of a skyrmion lattice (SkL) was first reported~\cite{Muehlbauer2009}, had been shown to posses a helimagnetic ground state in zero field~\cite{Ishikawa1976}, where the $q$-vectors of the energetically favoured helical modulations co-align with the magnetocrystalline easy axes, namely the cubic $\langle 100\rangle$-type axes~\cite{Nakanishi1980}. In FeGe (, another cubic chiral compound hosting Bloch SkL nearly up to room temperature~\cite{Yu2011}, the change of the cubic anisotropy on cooling was proposed to drive a reorientation of the helical $q$-vectors from $\langle 100\rangle$- to $\langle 111\rangle$-type directions~\cite{Ludgren1970, Siegfried2017}. In spite of these early observations and the increasing number of theoretical works~\cite{Skyrme1962, Pokrovsky1979, Bogdanov1989, Bogdanov1994, Roessler2006, Butenko2010, Buhrandt2013, Nagaosa2013, Leonov2015, Leonov2018, Leonov2019}, implying a clear impact of cubic anisotropy on helical spin states the effect of anisotropy on SkLs has only been recognized recently~\cite{Karube2016, Morikawa2017, Leonov2019}.

The cubic anisotropy was argued to play an essential role in the formation of a tilted conical and a SkL state at low temperatures in Cu$_2$OSeO$_3$~\cite{Chacon2018,Qian2018,Halder2018}, another cubic chiral magnet, where a higher-temperature SkL state was found earlier close to the Curie temperature ($T_\text C = 58\,K$)~\cite{Seki2012,Qian2016}. This low-temperature SkL forms when the tilted conical states gets destabilized above certain magnetic fields applied along one of the cubic $\langle 100\rangle$-type axes~\cite{Halder2018,Qian2018}. Elliptically distorted skyrmion strings were claimed to form in epitaxial thin films of MnSi in the presence of in-plane magnetic fields and out-of plane anisotropy~\cite{Wilson2012}.

\begin{figure}
	\includegraphics[width=\linewidth]{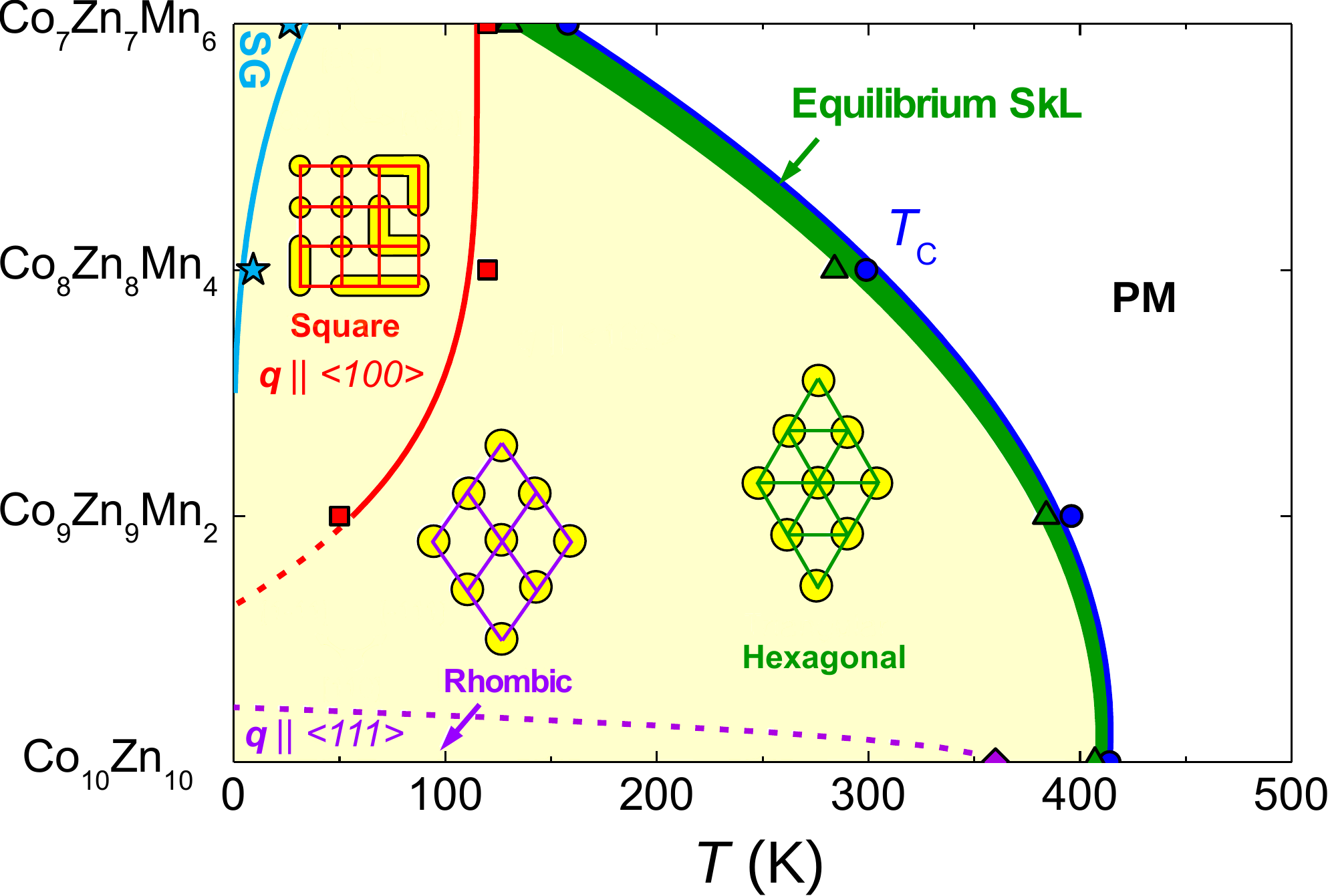}
	\caption{Schematic phase diagram of the equilibrium and metastable SkLs on the temperature-composition plane for four (Co$_{0.5}$Zn$_{0.5}$)$_{20-x}$Mn$_x$ compounds. Phase boundaries are reproduced from Ref.~\cite{Karube2020}. The green area indicates the thermodynamically stable SkL pocket below the Curie temperature (blue circles and line). The regime of the metastable SkL phases with pale yellow shading is separated into four regimes: 1) the transition from the equilibrium SkL into a metastable SkL ---preserving the hexagonal symmetry--- is indicated by green triangles, 2) the crossover from the hexagonal to a square metastable SkL phase in Co$_7$Zn$_7$Mn$_6$, Co$_8$Zn$_8$Mn$_4$ and Co$_9$Zn$_9$Mn$_2$ is indicated by red squares, 3) the transformation of the hexagonal into a rhombic metastable SkL state in Co$_{10}$Zn$_{10}$ is marked by a violet rhombus, 4) the onset of the reentrant spin-glass phase (SG) in Co$_7$Zn$_7$Mn$_6$ and Co$_8$Zn$_8$Mn$_4$ is marked by light blue stars. Characteristic temperatures have only been determined for the four compounds and the lines connecting the symbols are to support an easy visualization.}
	\label{phasediagram}
\end{figure}

Further prominent examples, where anisotropy affects the SkL state, have been reported for $\beta$-Mn-type (Co$_{0.5}$Zn$_{0.5}$)$_{20-x}$Mn$_x$ crystals~\cite{Hori2007}. These compounds belong to a new family of cubic chiral magnets, hosting skyrmions even well above room temperature~\cite{Tokunaga2015}. Similar to Fe$_x$Co$_{1-x}$Si~\cite{Muenzer2010,Milde2013}, their high-temperature SkL state can be quenched via field cooling to a metastable state that persists down to low temperatures and can be robust even against the reversal of magnetic field~\cite{Karube2016,Karube2017}. The symmetry of the metastable skyrmion lattice and the shape of the individual skyrmions show a large variation among these compounds, when changing the Co/Mn ratio, which implies a substantial role of cubic anisotropy in determining the internal structure of individual skyrmions and the lattice geometries of these metastable states~\cite{Morikawa2017}.
Figure\ref{phasediagram} gives an overview over the stable and metastable magnetic phases observed as a function of temperature and composition. Just below the Curie temperature a small pocket of SkL state emerges in all four compounds at moderate magnetic fields. Via field cooling this thermodynamically stable hexagonal SkL state can be quenched to lower temperatures. In Co$_7$Zn$_7$Mn$_6$, Co$_8$Zn$_8$Mn$_4$ and Co$_9$Zn$_9$Mn$_2$, this metastable SkL gradually transforms into a square lattice~\cite{Karube2017}. In thin lamellae, a rectangular TETRIS-like arrangement of elongated skyrmions is formed~\cite{Morikawa2017}. The elongation of individual skyrmions is likely caused by the increase in the length of the $q$-vectors with decreasing temperature and the conservation of the skyrmion number. The elongated skyrmions were found to co-align with the $\langle 100\rangle$-type cubic axes. On the other hand, in Co$_{10}$Zn$_{10}$ the metastable SkL becomes rhombic with two $q$-vectors pointing along the $\langle 111\rangle$-axes~\cite{Karube2020}. At the lowest temperatures, a spin glass state emerges in Co$_7$Zn$_7$Mn$_6$, Co$_8$Zn$_8$Mn$_4$, where the magnetic modulations are significantly disordered due to frustrated Mn spins~\cite{Karube2018}.

Apart from cubic chiral helimagnets, hosting Bloch-type skyrmions composed of helical spin modulations, skyrmions and antiskyrmions can form in crystals with axial symmetry. The lacunar spinels with a polar rhombohedral structure are the first compound family found to host N\'eel-type SkL composed of spin cycloids\cite{Kezsmarki2015}, while the antiskyrmion lattice state was first observed in Heusler compounds with non-centrosymmetric tetragonal structure~\cite{Nayak2017}. In contrast to the small phase pocket of thermodynamically stable SkLs in cubic chiral magnets, these N\'eel-type skyrmion and antiskyrmion lattices can be stable down to zero temperature~\cite{Bordacs2017, Leonov2017}. In these axially symmetric materials anisotropy can emerge in two forms: 1) the anisotropy of the Dzyaloshinskii-Moriya interaction (DMI) and 2) the uniaxial magnetocrystalline anisotropy. The former confines the magnetic modulation vectors to the plane perpendicular to the unique high-symmetry axis, thereby largely increasing the thermal stability range of the SkL state, while the second favours collinear states, hence suppressing modulations~\cite{Kezsmarki2015, Nayak2017, Ehlers2016, Bordacs2017, Leonov2017, Butykai2019, Geirhos2020, Gross2020}.

An omnifarious variety of exotic long-periodic magnetic patterns in addition to the well-known helical, conical and hexagonal SkL were observed in Co-Zn-Mn family, including square and rhombic skyrmion lattices \cite{Karube2016}, TETRIS-shaped skyrmions \cite{Morikawa2017}, skyrmion chains, meron-antimeron lattice, chiral soliton lattice \cite{Yu2018, Karube2020}, smectic liquid-crystalline arrangement of skyrmions \cite{Nagase2019} and domain-wall skyrmions \cite{Nagase2020}. Stability mechanisms of some of these textures and underlying physics in these materials, such as influence of spin disorder, frustration and magnetic anisotropy is still far from being understood. First of all, by adopting the simple model for chiral cubic B20-type magnets \cite{Maleyev2006} one can assume that the $q$-vector magnitude should only depend on the exchange stiffness $(J)$, DMI $(D)$ and weak anisotropic exchange constants. Indeed, micromagnetic simulations have shown, that the anisotropic deformation of the skyrmion lattice into a TETRIS-like pattern can be reproduced by the decrease of the $J/D$ ratio in the presence of cubic anisotropy \cite{Ukleev2019}. Strong easy-axis anisotropy is also suggested to stabilize domain-wall skyrmions \cite{Nagase2020} and the smectic skyrmion phase \cite{Nagase2019}. Therefore, a reliable measurement of the cubic anisotropy constants in CoZnMn is highly desired to correlate theories and experimental observations.

While these examples highlight the importance of magnetocrystalline anisotropy with regard to the stability, symmetry and shape of individual skyrmions and SkLs, surprisingly, quantitative and systematic experimental studies on the role of anisotropy in cubic skyrmion hosts are virtually non-existing. This provides the main motivation for the present work. Here, using ferromagnetic resonance (FMR) spectroscopy, we investigate the evolution of magnetocrystalline anisotropy in (Co$_{0.5}$Zn$_{0.5}$)$_{20-x}$Mn$_x$ compounds with composition ($x$=0,2,4,6) and temperature. Moreover, in comparison to the results of small angle neutron scattering (SANS) experiments, we analyze if changes in anisotropy can account for changes observed in the orientation and the length of the $q$-vectors and may eventually lead to transformations between different SkL arrangements. This material class serves as an ideal playground for such study, since the material composition can be tuned whilst keeping the cubic chiral symmetry, but vastly changing the ordering temperature, the magnetic properties and --as demonstrated by this work-- the strength and the nature of cubic anisotropy.

\section{Experimental methods}

In order to determine the magnetocrystalline anisotropy energy, we studied the angular dependence of the FMR field in the field-polarized ferromagnetic state of (Co$_{0.5}$Zn$_{0.5}$)$_{20-x}$Mn$_x$ with $x$=0,2,4,6 at temperatures ranging form 4\,K to 400\,K. The specimens were ${110}$-type cuts of single crystals polished to a cylindrical disc-form with a typical diameter of $\sim$1\,mm and thickness of $\sim$200\,$\mu$m. Field-swept FMR spectra were recorded at two different frequencies, X-band (9.4\,GHz) and Q-band (34\,GHz), in 5$^\circ$ steps upon rotation of the magnetic field in the $(1\bar{1}0)$ plane. Since the field polarized ferromagnetic state is achieved already at 1\,kOe in all the four compounds irrespective of the temperature, the FMR line, located always above 3\,kOe, does not overlap with either the collective modes of the modulated states or spectral features related to metamagnetic transitions.

\begin{figure}
	\includegraphics[width=\linewidth]{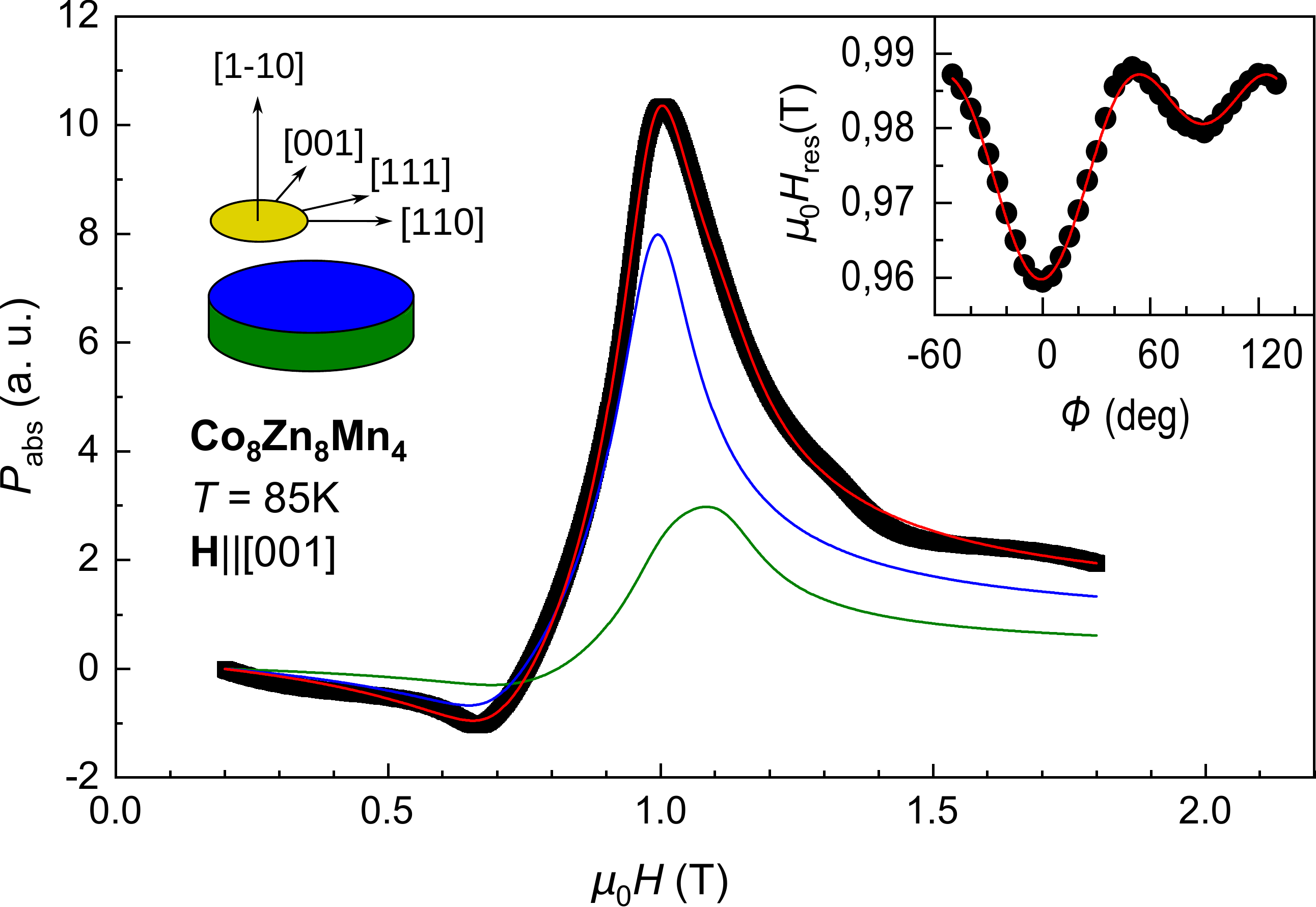}
	\caption{A field-swept FMR (microwave absorption) spectrum of Co$_8$Zn$_8$Mn$_4$ recorded at $\sim$85\,K with the magnetic field along the $[001]$ axis. The spectrum can be decomposed of two parts: A sharp resonance (blue line) originating from the top and bottom face of the sample and a broader peak (green line) coming from the mantle of the cylindrical disk, as sketched in the top-left corner. The dependence of the resonance field on $\Phi$ is shown in the inset, where $\Phi$ is the angle spanned by the field and the $[001]$ axis upon rotation of the magnetic field within the $(1\bar{1}0)$ plane. The field rotation plane, indicated as a pale yellow disk in the top left corner is parallel to the top and bottom surface of the sample.} \label{spectrum}
\end{figure}

The FMR spectra, see Fig.~\ref{spectrum} for a representative field-dependent microwave absorption spectrum, were reproduced and the resonance fields, $H_\text{res}$,  were determined using the Landau-Lifshitz-Gilbert equation~\cite{LandauLifshitz1935,Gilbert1955}
\begin{align}
	\frac{d\vec M}{dt} = -\gamma\vec M\times\vec H-\gamma\frac{\alpha}{M_\text s}\vec M\times \frac{d\vec M}{dt}\text, \label{landau}
\end{align}
where $\vec M$ is the time-dependent magnetization, $M_\text s$ the saturation magnetization, $\gamma$ the gyromagnetic ratio and $\vec H$ the applied magnetic field. Besides $H_\text{res}$, the Gilbert-damping parameter $\alpha$ and magnetization were also extracted from the fit. For all samples, the latter showed good agreement with magnetization data measured directly by SQUID.

Due to the skin effect characteristic to such metallic samples, the magnetization dynamics can be described by the thin-film variant of the Kittel formula~\cite{Kittel1947}
\begin{align}
	\frac{\omega}{\gamma} = H_\text{res,corr} = \sqrt{H_\text{res}\left[H_\text{res}+4\pi M\cdot\sin^2\left(\theta\right)\right]}\text, \label{cor}
\end{align}
where $\theta$ is the angle between the magnetization and the surface normal of the metallic specimen and $H_\text{res,corr}$ the resonance field corrected for the dynamic demagnetization effect. According to the relation above, only a thin surface layer of the metallic specimen contributes to the microwave absorption. Due to the geometry of the samples, the main contribution comes from the top and bottom surface of the cylinder. These are in-plane magnetized, thus, the main resonance peak is not affected by dynamic demagnetization. In contrast, the contribution from the surface around the mantle is subject to dynamic demagnetization and, thus, treated by integrating over the angle $\theta$ spanned by the surface normal and the magnetization. Since the dynamic demagnetization varies with $\theta$ around the mantle, the resonance field changes accordingly, leading to a second broader peak partially overlapping with the main sharp peak corresponding to the top and bottom layers. In addition, a dip at the left side of the main resonance is expected to appear, related to the so-called antiresonance~\cite{Kittel1951, Tannenwald1959, Morrish1965}. Its onset and width are controlled by the magnetization, while the Gilbert damping determines the linewidth of the main peak without capturing this dip.

Our analysis based on Eqs.~(1) and (2) could reproduce all features of the field-swept FMR spectra, as shown in Fig.~\ref{spectrum}, thus, facilitated a precise determination of $\alpha$ and $H_\text{res}(\Phi)$, where $\Phi$ measures the angle of the magnetic field with respect to the $[001]$ axis upon its rotation in the $(1\bar{1}0)$ plane. The angular dependence of $H_\text{res}$ is governed by the cubic anisotropy. We determined the cubic anisotropy parameters, $K_1$ and $K_2$ from a fitting routine based on the Smit-Suhl formula~\cite{SmitBeliers1955,Suhl1955}:

\begin{align}
	H_\text{res} = \frac{1}{m\sin\theta}\sqrt{\left.\frac{\partial^2\mathcal E}{\partial\theta^2}\right|_0\left.\frac{\partial^2 \mathcal E}{\partial\varphi^2}\right|_0 - \left(\left.\frac{\partial^2\mathcal E}{\partial\theta\partial\varphi}\right|_0\right)^2} \text,
\end{align}
where
\begin{align}
	\mathcal E = -\vec M\vec H + K_1\left(m_x^2m_y^2+m_x^2m_z^2+m_y^2m_z^2\right) +\nonumber \\ + K_2m_x^2m_y^2m_z^2 + \dots
\label{cubic_terms}
\end{align}
is the free energy, including the Zeeman term as well as fourth-order ($K_1$) and sixth-order($K_2$) cubic anisotropy terms with $m_x$, $m_y$ and $m_z$ being the direction cosines of the magnetization, while $\theta$ and $\Phi$ are the polar and azimuthal angle. Examples of $H_\text{res}(\Phi)$ curves and the corresponding fits are displayed in the inset of Fig.~\ref{spectrum} and in Figs.~\ref{3danisotropy}(a)-(c) for Co$_8$Zn$_8$Mn$_4$ at low temperature and for Co$_{10}$Zn$_{10}$ both at low and near room temperature.

In order to correlate changes of the anisotropy with transformations of the magnetic modulations, we also present SANS data for Co$_8$Zn$_8$Mn$_4$ and Co$_{10}$Zn$_{10}$ that are reproduced from previous works~\cite{Karube2016,Karube2020}. SANS images were recorded on the $(100)$-plane for Co$_8$Zn$_8$Mn$_4$ and on the $(110)$-plane for Co$_{10}$Zn$_{10}$. To determine the orientation of the favoured $q$-vectors, the angular dependence of the SANS intensity was fitted with Gaussian peaks. The width of peaks corresponding to symmetry-equivalent directions were averaged.
In addition, we also compare the temperature dependence of anisotropy parameters and $|q|$ values for all the four compounds, the latter reproduced from a previous SANS study~\cite{Karube2020}.

\section{Results and Discussion}

\subsection{Magnetocrystalline Anisotropy}

\begin{figure}
	\includegraphics[width=\linewidth]{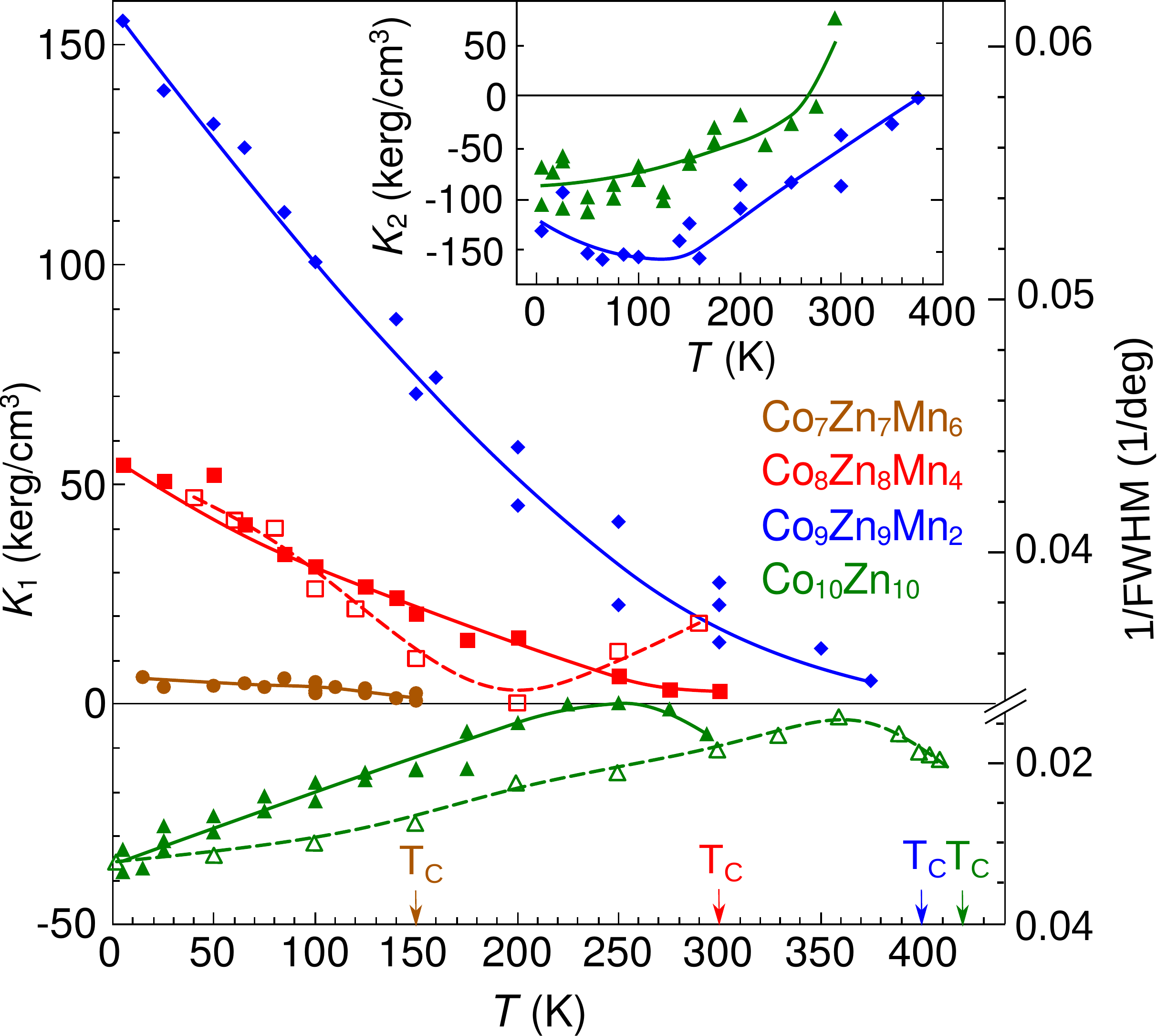}
	\caption{Temperature dependence of the cubic anisotropy constants for Co$_{10}$Zn$_{10}$ (green), Co$_9$Zn$_9$Mn$_2$ (blue), Co$_8$Zn$_8$Mn$_4$ (red) and Co$_7$Zn$_7$Mn$_6$ (brown). The corresponding critical temperatures of the Curie ordering are indicated by arrows. The anisotropy constant $K_1$ (left scale) is indicated by full symbols and with full lines as guides to the eye. The inverse of the full width at half maximum (FWHM) for the SANS peaks (right scale) are plotted by open symbols connected by dashed lines for Co$_{10}$Zn$_{10}$ (green) and Co$_8$Zn$_8$Mn$_4$ (red). The sixth order cubic anisotropy $K_2$ is plotted in the insert for Co$_{10}$Zn$_{10}$ (green) and Co$_9$Zn$_9$Mn$_2$ (blue).}
	\label{anisotropy}
\end{figure}

\begin{figure*}
	\includegraphics[width=\textwidth]{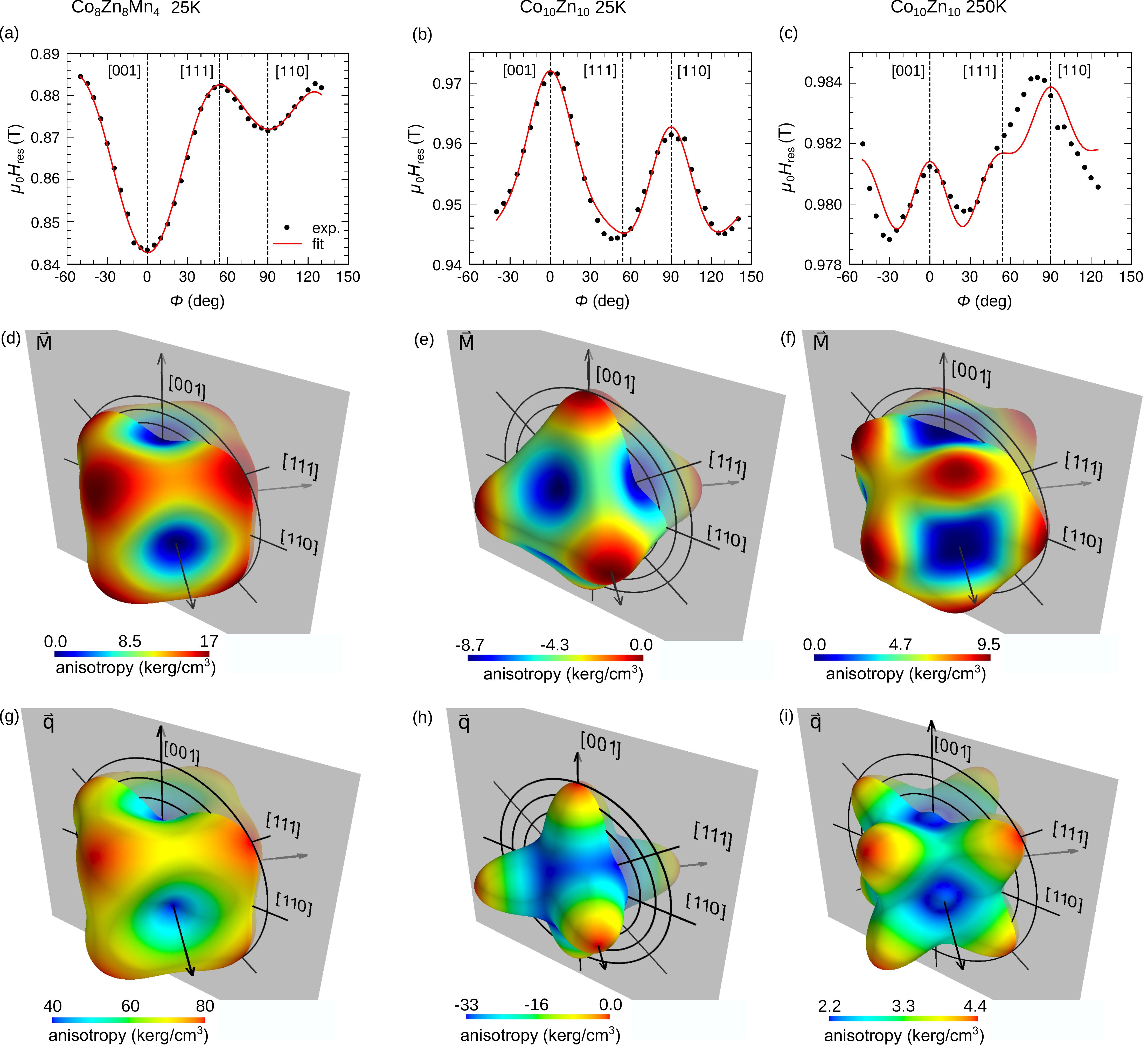}
	\caption{Angular dependence of the resonance field in the $(1\bar{1}0)$-plane is shown in panel (a) for Co$_8$Zn$_8$Mn$_4$ at 25\,K and in panels (b) and (c) for Co$_{10}$Zn$_{10}$ at 25\,K and 250\,K, respectively. Black circles are the experimental data points and red lines show the fit of the magnetocrystalline anisotropy using Eqs. (3) and (4). Panels (d)-(f) display the anisotropy energy surface of the ferromagnetic state, calculated based on the $K_1$, $K_2$, and $K_3$ values obtained from the fits in the corresponding panels above. Panels(g)-(i) show the anisotropy energy surface of the helical state depending on the orientation of the helical $q$-vector, again calculated using the parameters obtained in the corresponding top panels. Arrows in panels (d)-(i) indicate the $\langle 100\rangle$-type axes. In the rotation plane of the magnetic field, the grey $(1\bar{1}0)$ plane, $\langle 111\rangle$-type, and $\langle 110\rangle$-type axes are also shown.}
	\label{3danisotropy}
\end{figure*}

Figure \ref{anisotropy} summarizes the temperature dependence of the cubic anisotropy parameters for all the four compounds. As a common trend, the anisotropy energy decreases with increasing temperature and vanishes smoothly around the Curie temperature, except in Co$_{10}$Zn$_{10}$. Besides its strong thermal variation, the anisotropy is sensitive to the Co/Mn ratio and drastically enhanced with increasing Co content. At low temperatures Co$_8$Zn$_8$Mn$_4$ and Co$_9$Zn$_9$Mn$_2$ are respectively characterized by $\sim$10 and $\sim$30 times larger fourth-order cubic anisotropy ($K_1$) than Co$_7$Zn$_7$Mn$_6$. Co$_{10}$Zn$_{10}$ does not follow this tendency, instead the $K_1$ anisotropy term changes sign in this compound. Moreover, with increasing Co content the sixth-order cubic anisotropy ($K_2$) becomes relevant: While it is negligible in Co$_7$Zn$_7$Mn$_6$ and Co$_8$Zn$_8$Mn$_4$, in Co$_9$Zn$_9$Mn$_2$ it is comparable to $K_1$ and becomes even the dominant term in Co$_{10}$Zn$_{10}$. In contrast to the strong composition dependence of the anisotropy, the saturation magnetization per magnetic ion varies by not more than 20\,\% among the four compounds. In terms of the temperature dependence of the anisotropy, Co$_{10}$Zn$_{10}$ is again an outlier: $K_1$ approaches zero around 250\,K, i.e. well below the Curie temperature, where the $K_2$ term changes sign.

When the FM state emerges in a cubic crystal, the favoured direction of the magnetic moment is determined by the cubic anisotropy terms given in Eq.~\ref{cubic_terms}, in case the magnetically induced structural distortion is negligible. If $K_1$ is the dominant among the cubic anisotropy terms, the easy axes of the magnetization is along the $\langle 100\rangle$-type directions for $K_1>0$ and along the $\langle 111\rangle$-type directions for $K_1<0$. Similarly, when $K_2$ governs the anisotropy energy, the easy axes are of $\langle 100\rangle$-type for $K_2>0$ and $\langle 111\rangle$-type for $K_2<0$. (When $K_1$ and $K_2$ have opposite signs the easy axes of magnetization can even point along the $\langle 110\rangle$-type directions.) For a more complex magnetic texture, like the helical spin order, the magnetic anisotropy energy has to be evaluated for the whole magnetic unit cell, i.e. by summing up for all the spins in one period of the helix. Since the $q$-vector is perpendicular to the spin rotation of the helix, the energetically favoured directions of the $q$-vectors can be determined if the values of $K_1$ and $K_2$ are known~\cite{Chacon2018,Halder2018}, as will be shown later.

Correspondingly, Co$_7$Zn$_7$Mn$_6$, Co$_8$Zn$_8$Mn$_4$ and Co$_9$Zn$_9$Mn$_2$ are characterized by easy axes of the magnetization along the $\langle 100\rangle$-type directions at all temperatures, though the anisotropy energy is considerably lowered upon approaching the Curie temperature. As an example, the anisotropy energy surface in the FM state of Co$_8$Zn$_8$Mn$_4$ is shown in Fig.~\ref{3danisotropy}(d), as calculated using $K_1$ and $K_2$ values obtained from the fit of $H_\text{res}(\Phi)$ in Fig.~\ref{3danisotropy}(a). This energy surface represents the anisotropy of the FM state or equivalently the anisotropy experienced by an individual spin.

For a helix, where the spins rotate in the plane perpendicular to the $q$-vector, the energetically favoured orientation of the $q$-vector is such that the integral of the anisotropy energy is minimal around the closed loop in the plane perpendicular to the $q$-vector. (In this simple approach we treat the magnetic modulations as fully harmonic spin helices and disregard possible effects of anharmonicity.) When the easy axes of magnetization are the $\langle 100\rangle$-type directions, the preferred $q$-vectors also point along the $\langle 100\rangle$-type axes, as can be easily followed in Fig.~\ref{3danisotropy}(d). This can be further checked in Fig.~\ref{3danisotropy}(g), which displays the anisotropy energy versus the orientation of the helical $q$-vectors for Co$_8$Zn$_8$Mn$_4$ at low temperature. Qualitatively the same scenario applies also for Co$_7$Zn$_7$Mn$_6$ and Co$_9$Zn$_9$Mn$_2$. However, the anisotropy in Co$_7$Zn$_7$Mn$_6$ is considerably weaker, thus, the anisotropy energy surface is more spherical with shallow minima along $\langle 100\rangle$-type directions. In contrary, the anisotropy energy surface of Co$_9$Zn$_9$Mn$_2$ has nearly a cuboid shape due to large $K_1$ and $K_2$ values with $K_1/K_2$ close to unity.

In contrast, in Co$_{10}$Zn$_{10}$ with $K_1<0$ and $K_2<0$ at low temperatures, the easy axes of the magnetization point along the $\langle 111\rangle$-type directions, as clear from Fig.~\ref{3danisotropy}(e). While the helical modulation vectors favour the same $\langle 111\rangle$-type directions, $q$-vectors between $\langle 111\rangle$- and $\langle 110\rangle$-type axes have nearly the same energy and only the $\langle 100\rangle$-type axes are clearly distinguished as unfavoured directions, as can be traced in Fig.~\ref{3danisotropy}(h). Moreover, with increasing temperature both anisotropy terms weaken strongly and $K_2$ even changes sign between 200\,K and 250\,K, temperatures far below $T_C=420$\,K. Consequently, the $\langle 100\rangle$-type directions, being the hard axes of magnetization at low-temperatures, nearly become the easy axes [Fig.~\ref{3danisotropy}(f)]. In this temperature range the weak positive eighth-order anisotropy term ($K_3\approx 23\,\text{kerg}$) also becomes relevant in defining the location of the shallow energy minima. In fact, at these temperatures the favoured $q$-vectors do not point along any of the high-symmetry directions, instead they are located around circles centered at $\langle 100\rangle$-type axes. A Mexican hat-like form of the energy surface in that region can be seen in Fig.~\ref{3danisotropy}(i).

\begin{figure*}[t]
	\includegraphics[width=\textwidth]{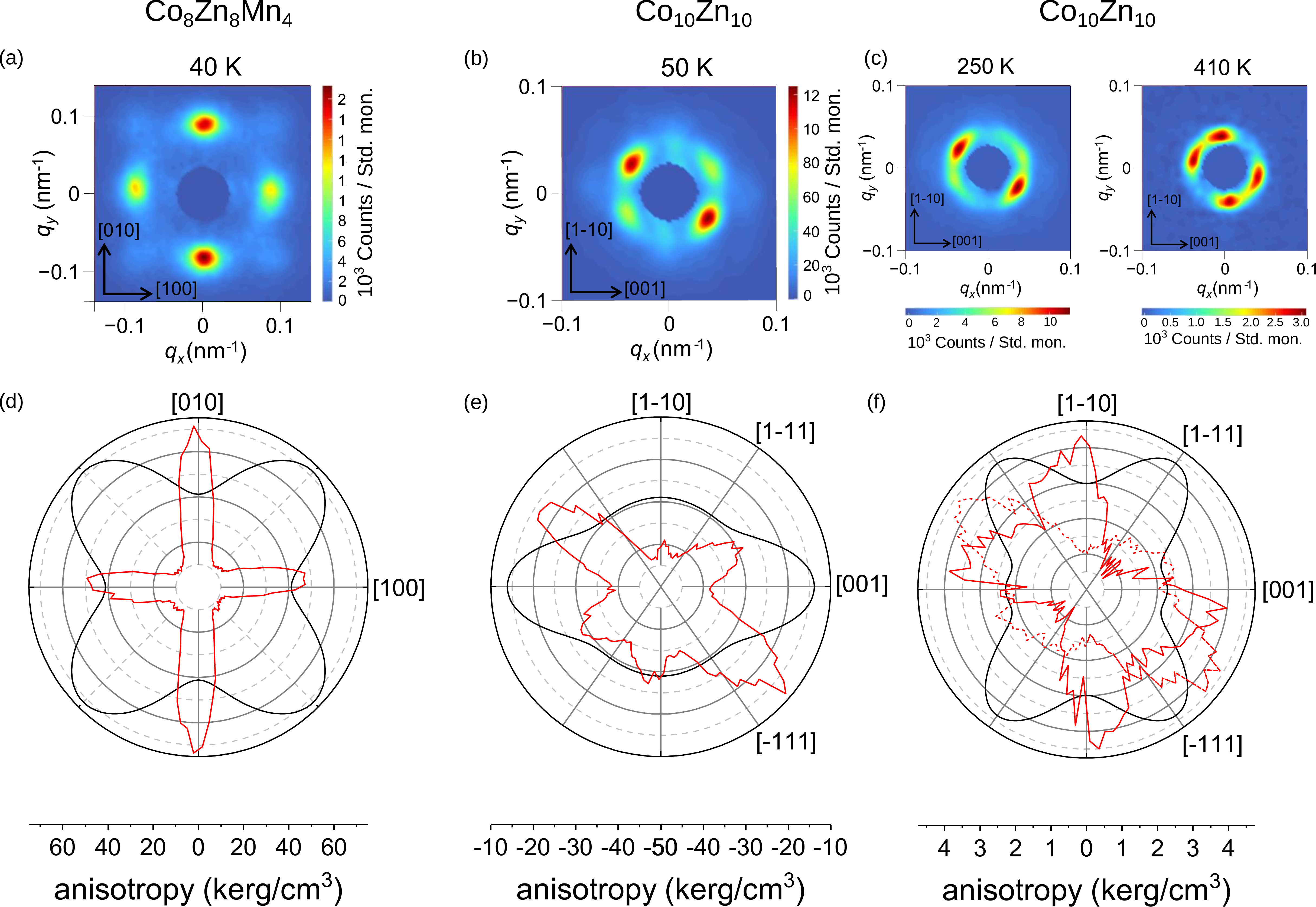}
	\caption{(a) SANS intensity of Co$_8$Zn$_8$Mn$_4$ recorded on the $(001)$ plane at 40\,K. SANS intensity of Co$_{10}$Zn$_{10}$ on the $(110)$ plane at (b) 50\,K and (c) 250\,K and 410\,K. Direct comparison between the dependence of the magnetocrystalline anisotropy energy of the helix (black curves) and the SANS intensity (red curves) on the direction of the $q$-vector, in polar-plot representation, for Co$_8$Zn$_8$Mn$_4$ at 40\,K (d) and Co$_{10}$Zn$_{10}$ (b) at 50\,K as well as (c) 250\,K (solid curve) and 410\,K (dashed curve). In panels (d)-(f), the anisotropy energy curves are the corresponding cross-sections of the anisotropy energy surfaces in Figs.~\ref{3danisotropy}(g)-(i), while the azimuthal angular-dependent SANS intensity curves are obained by radial averaging of the SANS images in panels (a)-(c).}
	\label{qvectoranisotropy}
\end{figure*}

\subsection{Magnetic Modulations and Anisotropy}

In what follows, we perform a comparative analysis of FMR and SANS data to reveal how the variation of cubic anisotropy with temperature and composition affects the magnetic modulations in this series of (Co$_{0.5}$Zn$_{0.5}$)$_{20-x}$Mn$_x$ compounds. The SANS data are reproduced from Refs.~\cite{Karube2016,Karube2020} and re-analyzed here to gain more insight into the azimuthal angular distribution of the $q$-vectors. At low temperatures the SANS intensity is confined to the $\langle 100\rangle$-type directions in Co$_9$Zn$_9$Mn$_2$, Co$_8$Zn$_8$Mn$_4$ and Co$_7$Zn$_7$Mn$_6$, i.e. the helical $q$-vectors favour these directions, as shown in the SANS image in Fig.~\ref{qvectoranisotropy}(a), recorded on the $(001)$ plane for Co$_8$Zn$_8$Mn$_4$. In contrast, in the SANS intensity of Co$_{10}$Zn$_{10}$, displayed over the $(110)$ plane in Fig.~\ref{qvectoranisotropy}(b), the maxima are located at the $\langle 111\rangle$ directions at low temperatures. In both materials, the difference in the intensity of symmetry-equivalent $q$-vectors originates from unequal population of the corresponding helical domains.

The azimuthal angular distribution of the $q$-vectors becomes broader with increasing temperature in both materials, as clear from the increase of the full width at half maximum (FWHM) for the SANS maxima, plotted in Fig.~\ref{anisotropy}. While in Co$_{10}$Zn$_{10}$ the FWHM is more than doubled ($~$31$^\circ \rightarrow ~$71$^\circ$) between 5\,K and 370\,K, it is only increased by about 20\% ($~$23$^\circ \rightarrow ~$29$^\circ$) in Co$_{8}$Zn$_{8}$Mn$_4$. In fact, the inverse of the FWHM follows a temperature dependence similar to that of the $K_1$ values with a small difference close to the Curie temperatures. This shows that the strength of the anisotropy directly controls the width of the azimuthal angular distribution of the $q$-vectors. In addition to the enhanced orientational disorder of the $q$-vectors towards higher temperatures, in Co$_{10}$Zn$_{10}$ the position of the favoured $q$-vectors moves away from the $\langle 111\rangle$-type to $\langle 100\rangle$-type directions. This happens between 250\,K and 410\,K, as can be traced well in Figs.~\ref{qvectoranisotropy}(c) and (d).

Since apparently the helical $q$-vectors arrange in a way to minimize the anisotropy energy, in Figs.~\ref{qvectoranisotropy}(d)-(f) we directly compare the azimuthal angular dependence of the SANS intensity and the anisotropy energy of spin helices depending on the orientation of their $q$-vectors. In Co$_{8}$Zn$_{8}$Mn$_4$ the anisotropy energy of the spin helix has minima along the $\langle 100\rangle$-type axes, coinciding with the SANS maxima. In Co$_{10}$Zn$_{10}$ a similar confinement of the $q$-vectors by anisotropy is observed around the $\langle 111\rangle$-type axes at low temperature, though the width of the SANS maxima are considerably broader due to the weaker anisotropy in this compound [Fig.~\ref{qvectoranisotropy}(e)]. Note that the anisotropy energy of the spin helix shows little dependence on the orientation of the $q$-vectors in the range of $[1\bar{1}\bar{1}]$--$[1\bar{1}0]$--$[1\bar{1}1]$. At 250\,K the dependence of the anisotropy on the orientation of the helical $q$-vectors is getting even weaker with $\Delta\mathcal E\approx 2$\,kerg/cm$^3$, compared to $\Delta\mathcal E\approx 15$\,kerg/cm$^3$ at 50\,K. Consequently, though the global minima are located around circles centered at $\langle 100\rangle$-type axes, as shown in Fig.~\ref{3danisotropy}(i), the helical $q$-vectors are easily trapped at local minima, as clear from Fig.~\ref{qvectoranisotropy}(f).

\subsection{Magnetic Modulations and Gilbert Damping}

\begin{figure*}
	\includegraphics[width=\textwidth]{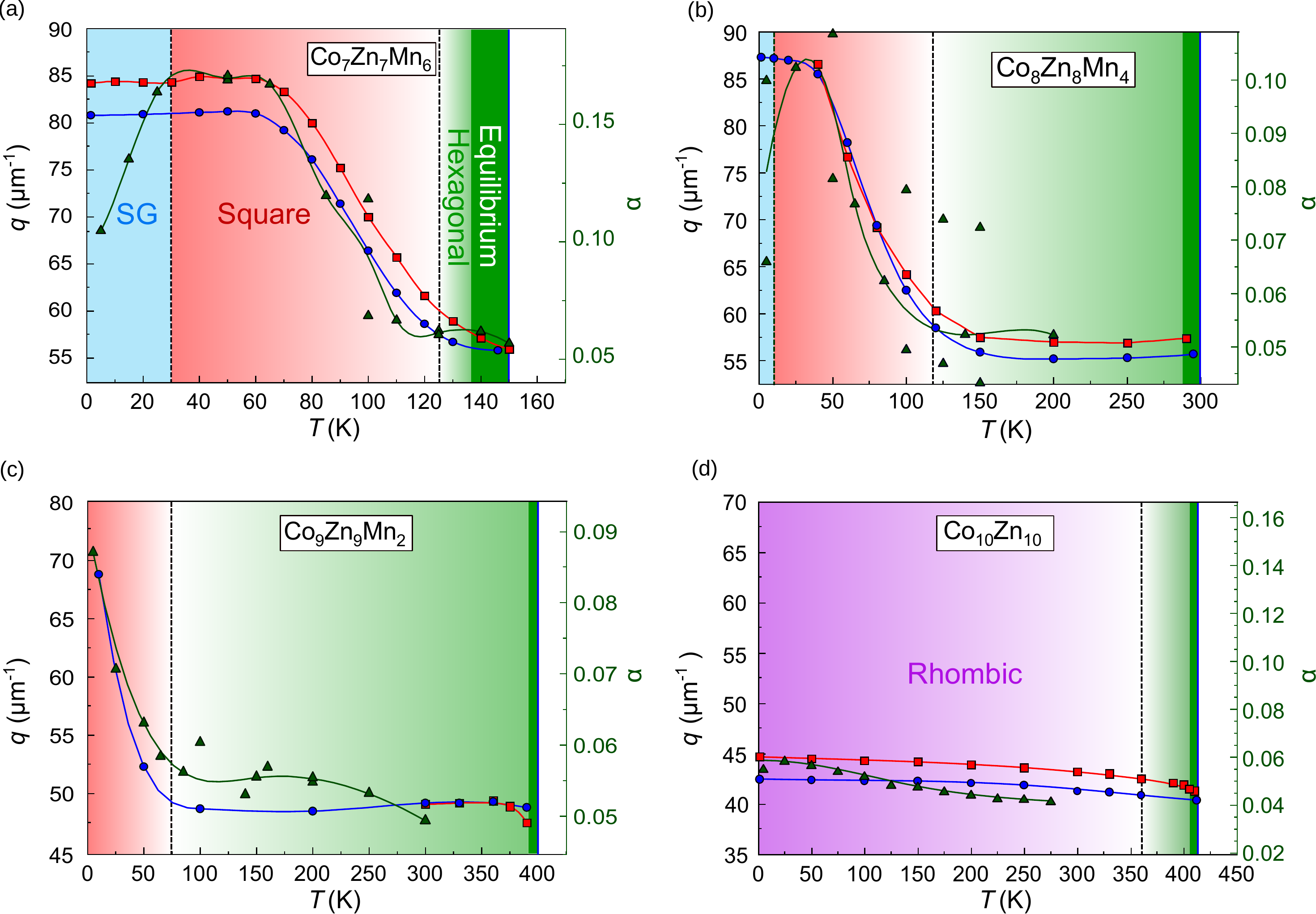}
	\caption{Comparison between the temperature dependence of the length of the $q$-vectors (left scale) and the Gilbert-damping parameter $\alpha$  (right scale) in the metastable SkLs of (a) Co$_7$Zn$_7$Mn$_6$, (b) Co$_8$Zn$_8$Mn$_4$, (c) Co$_9$Zn$_9$Mn$_2$ and (d) Co$_{10}$Zn$_{10}$. Blue circles show the $q$-vector length in the SkL states and red squares in the helical phase, while green triangles indicate $\alpha$. The green area represents the equilibirium SkL state just below the Curie temperature (blue line). The green shaded area represents the regime over which the metastable SkL preserves the hexagonal symmetry, while the red and violet shaded areas indicate the square and rhombic SkL states, respectively. The light blue shaded area stands for the reentrant spin-glass phase. (see also Fig.~\ref{phasediagram})}
	\label{qandalpha}
\end{figure*}

At the Curie temperature the Gilbert-damping parameter $\alpha$ lies between 0.04 and 0.05 in these compounds, which is a typical value for ferromagnetic Cobalt. Upon cooling $\alpha$ is enhanced in all the four compounds, however, the strength of the increase depends strongly on the Co/Mn ratio, as clear from Fig.~\ref{qandalpha}. In Co$_{10}$Zn$_{10}$ $\alpha$ increases only by $\sim$50\,\% while it is nearly tripled in Co$_{7}$Zn$_{7}$Mn$_6$. Surprisingly, $\alpha$ and the length of the $q$-vectors show quantitatively similar thermal variations in all compounds except for Co$_{10}$Zn$_{10}$: They are both constant through the thermodynamically stable as well as metastable hexagonal SkL phases but exhibit a sudden uprise at the transformation from the metastable hexagonal to square lattice (TETRIS-like) phase. On the other hand, $\alpha$ and the width of the $q$-vectors do not show a similar temperature dependence. Instead, the temperature dependence of the latter is reminiscent of that of $K_1$, as seen in Fig.\,\ref{anisotropy}. Although we do not have a microscopic picture capturing the apparent correlation between the Gilbert damping and the length of the $q$-vectors, their strikingly similar temperature dependences point to a common underlying microscopic mechanism. One possibility is that the enhancement of magnetic anisotropy towards low temperatures is responsible for, although not directly proportional to, the increase observed both in the spin relaxation and the anharmonicity. Another possible scenario is that antiferromagnetic correlations of Mn spins developing towards low temperatures weaken the ferromagnetic exchange but not the DMI between Co spins, and hence increase the length of the $q$-vectors. This would also explain why the increase of $|q|$ starts at a lower temperature in Co$_9$Zn$_9$Mn$_2$ with higher Co content  than in Co$_8$Zn$_8$Mn$_4$ and Co$_7$Zn$_7$Mn$_6$, although $K_1$ is larger in Co$_9$Zn$_9$Mn$_2$ than in Co$_8$Zn$_8$Mn$_4$ and Co$_7$Zn$_7$Mn$_6$. In this picture the low-temperature enhancement of Gilbert damping is governed by static or dynamic disorder of Mn spins, since the helimagnetic order is suppressed and the Co spins are fully spin polarized in magnetic fields above 1\,T, where $\alpha$ was determined from the FMR spectra. For compounds with a SG ground state, a drop in $\alpha$ is found below the onset of the spin-glass phase, as observed previously in canonical spin-glass materials~\cite{Ford1976}. Again Co$_{10}$Zn$_{10}$ is an outlier where both $\alpha$ and $|q|$ exhibit little temperature dependence with no clear correlation between the two.

\section{Conclusions}

In the present work we studied the evolution of the cubic anisotropy in a series of (Co$_{0.5}$Zn$_{0.5}$)$_{20-x}$Mn$_x$ cubic chiral magnets for $x=0,2,4,6$. While these compounds are isostructural and the magnitude of the saturated magnetic moment does not vary by more than 20\% upon changing the Co/Mn ratio, the magnetic interactions show a strong dependence on the composition, as already reflected by the large variation of the Curie temperature between 150\,K and 410\,K. Here we found that this series offers an excellent laboratory to study the effects of anisotropy on chiral spin textures. This is because not only the magnitude of the anisotropy shows drastic variation with the Co/Mn ratio and the temperature but also the character of the anisotropy changes, as reflected by the reorientation of the easy axes from $\langle 100\rangle$-type to $\langle 111\rangle$-type directions with increasing Co content. Due to the close competition of the fourth and sixth order cubic anisotropy terms, the eights order anisotropy also comes into play in Co$_{10}$Zn$_{10}$ near room temperature. As a consequence, the helical $q$-vectors experience a Mexican hat-like potential centered around the $\langle 100\rangle$ directions. In all these compounds the anisotropy plays the most obvious role in governing the orientation of the helical $q$-vectors and the lattice geomotry in the metastable SkL states. While anisotropy clearly affects the orientation of helical $q$-vectors and may induced anharmonicity in the magnetic modulations, the high-temperature thermodynamically stable SkL states in all these compounds skyrmions still form regular hexagonal lattices with a 120$^\circ$ triple-$q$ structure. On the other hand, at low temperatures, when the anisotropy gets enhanced, it plays a key role in the metastable SkL states by reducing the originally cylindrical symmetry of individual skyrmions. The elongation of the skyrmions then leads to transformation of the hexagonal lattice to rhombohedral form in Co$_{10}$Zn$_{10}$ and to square form in the other three compounds, as dictated by the favoured $q$-vectors. The Gilbert damping turns out to be also a good indicator of transformations within the metastable SkLs. While its value is comparable to that of the elemental Co in the paramagnetic state and also below the Curie temperature of these compounds, it is highly increased at low temperatures where the metastable SkL states exhibit distortions with respect to the hexagonal structure.

\section{Acknowledgment}

We would like to express our special thanks to Vladimir Tsurkan for polishing the cylindric speciemens, to Dana Vieweg, who performed the magnetization measurements on our samples, and Veronika Fritsch for the Laue orientation imaging of the specimens. This research was supported by the German Research Foundation (DFG) via the Priority Program SPP2137, Skyrmionics,
under Grant Nos. KE 2370/1-1, EN 434/40-1

\bibliography{mybib.bib}{}

\end{document}